


\catcode`\@=11
\expandafter\ifx\csname @iasmacros\endcsname\relax
	\global\let\@iasmacros=\par
\else	\endinput
\fi
\catcode`\@=12


\def\rmb{\seventeenrm}

\def\singlespace{\baselineskip=\normalbaselineskip}


\def\nonarrower{\advance\leftskip by-\parindent
	\advance\rightskip by-\parindent}


\def\undertext#1{$\underline{\smash{\hbox{#1}}}$}
\def\boxit#1{\vbox{\hrule\hbox{\vrule\kern3pt
	\vbox{\kern3pt#1\kern3pt}\kern3pt\vrule}\hrule}}

\def\hence{\leavevmode\hbox{\bf .\raise5.5pt\hbox{.}.} }

\def\dalemb#1#2{{\vbox{\hrule height.#2pt
	\hbox{\vrule width.#2pt height#1pt \kern#1pt \vrule width.#2pt}
	\hrule height.#2pt}}}
\def\gtorder{\mathrel{\raise.3ex\hbox{$>$}\mkern-14mu
             \lower0.6ex\hbox{$\sim$}}}
\def\ltorder{\mathrel{\raise.3ex\hbox{$<$}\mkern-14mu
             \lower0.6ex\hbox{$\sim$}}}

\newdimen\fullhsize
\newbox\leftcolumn
\def\twoup{\hoffset=-.5in \voffset=-.25in
  \hsize=4.75in \fullhsize=10in \vsize=6.9in
  \def\fullline{\hbox to\fullhsize}
  \let\lr=L
  \output={\if L\lr
        \global\setbox\leftcolumn=\columnbox\global\let\lr=R \advancepageno
      \else \doubleformat \global\let\lr=L\fi
    \ifnum\outputpenalty>-20000 \else\dosupereject\fi}
  \def\doubleformat{\shipout\vbox{
    \fullline{\box\leftcolumn\hfil\columnbox}\advancepageno}}
  \def\columnbox{\leftline{\vbox{\makeheadline\pagebody\makefootline}}}
  \tolerance=1000 }

\catcode`\@=11					



\font\fiverm=cmr5				
\font\fivemi=cmmi5				
\font\fivesy=cmsy5				
\font\fivebf=cmbx5				

\skewchar\fivemi='177
\skewchar\fivesy='60


\font\sixrm=cmr6				
\font\sixi=cmmi6				
\font\sixsy=cmsy6				
\font\sixbf=cmbx6				

\skewchar\sixi='177
\skewchar\sixsy='60


\font\sevenrm=cmr7				
\font\seveni=cmmi7				
\font\sevensy=cmsy7				
\font\sevenit=cmti7				
\font\sevenbf=cmbx7				

\skewchar\seveni='177
\skewchar\sevensy='60


\font\eightrm=cmr8				
\font\eighti=cmmi8				
\font\eightsy=cmsy8				
\font\eightit=cmti8				
\font\eightbf=cmbx8				

\skewchar\eighti='177
\skewchar\eightsy='60


\font\ninei=cmmi9
\font\ninesy=cmsy9

\skewchar\ninei='177
\skewchar\ninesy='60


\font\tenrm=cmr10				
\font\teni=cmmi10				
\font\tensy=cmsy10				
\font\tenex=cmex10				
\font\tenit=cmti10				
\font\tensl=cmsl10				
\font\tenbf=cmbx10				
\font\tentt=cmtt10				
\font\tenss=cmss10				
\font\tensc=cmcsc10				
\font\tenbi=cmmib10				

\skewchar\teni='177
\skewchar\tenbi='177
\skewchar\tensy='60

\def\tenpoint{\ifmmode\err@badsizechange\else
	\textfont0=\tenrm \scriptfont0=\sevenrm \scriptscriptfont0=\fiverm
	\textfont1=\teni  \scriptfont1=\seveni  \scriptscriptfont1=\fivemi
	\textfont2=\tensy \scriptfont2=\sevensy \scriptscriptfont2=\fivesy
	\textfont3=\tenex \scriptfont3=\tenex   \scriptscriptfont3=\tenex
	\textfont4=\tenit \scriptfont4=\sevenit \scriptscriptfont4=\sevenit
	\textfont5=\tensl
	\textfont6=\tenbf \scriptfont6=\sevenbf \scriptscriptfont6=\fivebf
	\textfont7=\tentt
	\textfont8=\tenbi \scriptfont8=\seveni  \scriptscriptfont8=\fivemi
	\def\rm{\tenrm\fam=0 }%
	\def\it{\tenit\fam=4 }%
	\def\sl{\tensl\fam=5 }%
	\def\bf{\tenbf\fam=6 }%
	\def\tt{\tentt\fam=7 }%
	\def\ss{\tenss}%
	\def\sc{\tensc}%
	\def\bmit{\fam=8 }%
	\rm\setparameters\setbaselines\fi}


\font\twelverm=cmr12				
\font\twelvei=cmmi12				
\font\twelvesy=cmsy10	scaled\magstep1		
\font\twelveex=cmex10	scaled\magstep1		
\font\twelveit=cmti12				
\font\twelvesl=cmsl12				
\font\twelvebf=cmbx12				
\font\twelvett=cmtt12				
\font\twelvess=cmss12				
\font\twelvesc=cmcsc10	scaled\magstep1		
\font\twelvebi=cmmib10	scaled\magstep1		

\skewchar\twelvei='177
\skewchar\twelvebi='177
\skewchar\twelvesy='60

\def\twelvepoint{\ifmmode\err@badsizechange\else
	\textfont0=\twelverm \scriptfont0=\eightrm \scriptscriptfont0=\sixrm
	\textfont1=\twelvei  \scriptfont1=\eighti  \scriptscriptfont1=\sixi
	\textfont2=\twelvesy \scriptfont2=\eightsy \scriptscriptfont2=\sixsy
	\textfont3=\twelveex \scriptfont3=\tenex   \scriptscriptfont3=\tenex
	\textfont4=\twelveit \scriptfont4=\eightit \scriptscriptfont4=\sevenit
	\textfont5=\twelvesl
	\textfont6=\twelvebf \scriptfont6=\eightbf \scriptscriptfont6=\sixbf
	\textfont7=\twelvett
	\textfont8=\twelvebi \scriptfont8=\eighti  \scriptscriptfont8=\sixi
	\def\rm{\twelverm\fam=0 }%
	\def\it{\twelveit\fam=4 }%
	\def\sl{\twelvesl\fam=5 }%
	\def\bf{\twelvebf\fam=6 }%
	\def\tt{\twelvett\fam=7 }%
	\def\ss{\twelvess}%
	\def\sc{\twelvesc}%
	\def\bmit{\fam=8 }%
	\rm\setparameters\setbaselines\fi}


\font\fourteenrm=cmr10	scaled\magstep2		
\font\fourteeni=cmmi10	scaled\magstep2		
\font\fourteensy=cmsy10	scaled\magstep2		
\font\fourteenex=cmex10	scaled\magstep2		
\font\fourteenit=cmti10	scaled\magstep2		
\font\fourteensl=cmsl10	scaled\magstep2		
\font\fourteenbf=cmbx10	scaled\magstep2		
\font\fourteentt=cmtt10	scaled\magstep2		
\font\fourteenss=cmss10	scaled\magstep2		
\font\fourteensc=cmcsc10 scaled\magstep2	
\font\fourteenbi=cmmib10 scaled\magstep2	

\skewchar\fourteeni='177
\skewchar\fourteenbi='177
\skewchar\fourteensy='60

\def\fourteenpoint{\ifmmode\err@badsizechange\else
	\textfont0=\fourteenrm \scriptfont0=\tenrm \scriptscriptfont0=\sevenrm
	\textfont1=\fourteeni  \scriptfont1=\teni  \scriptscriptfont1=\seveni
	\textfont2=\fourteensy \scriptfont2=\tensy \scriptscriptfont2=\sevensy
	\textfont3=\fourteenex \scriptfont3=\tenex \scriptscriptfont3=\tenex
	\textfont4=\fourteenit \scriptfont4=\tenit \scriptscriptfont4=\sevenit
	\textfont5=\fourteensl
	\textfont6=\fourteenbf \scriptfont6=\tenbf \scriptscriptfont6=\sevenbf
	\textfont7=\fourteentt
	\textfont8=\fourteenbi \scriptfont8=\tenbi \scriptscriptfont8=\seveni
	\def\rm{\fourteenrm\fam=0 }%
	\def\it{\fourteenit\fam=4 }%
	\def\sl{\fourteensl\fam=5 }%
	\def\bf{\fourteenbf\fam=6 }%
	\def\tt{\fourteentt\fam=7}%
	\def\ss{\fourteenss}%
	\def\sc{\fourteensc}%
	\def\bmit{\fam=8 }%
	\rm\setparameters\setbaselines\fi}


\font\seventeenrm=cmr10 scaled\magstep3		


\newdimen\rp@
\newcount\@basestretchnum
\newskip\@baseskip
\newskip\headskip
\newskip\footskip


\def\setparameters{\rp@=.1em
	\headskip=24\rp@
	\footskip=\headskip
	\delimitershortfall=5\rp@
	\nulldelimiterspace=1.2\rp@
	\scriptspace=0.5\rp@
	\abovedisplayskip=10\rp@ plus3\rp@ minus5\rp@
	\belowdisplayskip=10\rp@ plus3\rp@ minus5\rp@
	\abovedisplayshortskip=5\rp@ plus2\rp@ minus4\rp@
	\belowdisplayshortskip=10\rp@ plus3\rp@ minus5\rp@
	\normallineskip=\rp@
	\lineskip=\normallineskip
	\normallineskiplimit=0pt
	\lineskiplimit=\normallineskiplimit
	\jot=3\rp@
	\setbox0=\hbox{\the\textfont3 B}\p@renwd=\wd0
	\skip\footins=12\rp@ plus3\rp@ minus3\rp@
	\skip\topins=0pt plus0pt minus0pt}


\def\setbaselines{\maxdepth=4\rp@\baselinestretch=\@basestretchnum}


\def\baselinestretch{\afterassignment\@basestretch\@basestretchnum}
\def\@basestretch{%
	\@baseskip=12\rp@ \divide\@baseskip by1000
	\normalbaselineskip=\@basestretchnum\@baseskip
	\baselineskip=\normalbaselineskip
	\bigskipamount=\the\baselineskip
		plus.25\baselineskip minus.25\baselineskip
	\medskipamount=.5\baselineskip
		plus.125\baselineskip minus.125\baselineskip
	\smallskipamount=.25\baselineskip
		plus.0625\baselineskip minus.0625\baselineskip
	\setbox\strutbox=\hbox{\vrule height.708\baselineskip
		depth.292\baselineskip width0pt }}



\def\makeheadline{\vbox to0pt{\baselinestretch=1000
	\vskip-\headskip \vskip1.5pt
	\line{\vbox to\ht\strutbox{}\the\headline}\vss}\nointerlineskip}

\def\makefootline{\baselineskip=\footskip\line{\the\footline}}

\def\big#1{{\hbox{$\left#1\vbox to8.5\rp@ {}\right.\n@space$}}}
\def\Big#1{{\hbox{$\left#1\vbox to11.5\rp@ {}\right.\n@space$}}}
\def\bigg#1{{\hbox{$\left#1\vbox to14.5\rp@ {}\right.\n@space$}}}
\def\Bigg#1{{\hbox{$\left#1\vbox to17.5\rp@ {}\right.\n@space$}}}


\mathchardef\alpha="710B
\mathchardef\beta="710C
\mathchardef\gamma="710D
\mathchardef\delta="710E
\mathchardef\epsilon="710F
\mathchardef\zeta="7110
\mathchardef\eta="7111
\mathchardef\theta="7112
\mathchardef\iota="7113
\mathchardef\kappa="7114
\mathchardef\lambda="7115
\mathchardef\mu="7116
\mathchardef\nu="7117
\mathchardef\xi="7118
\mathchardef\pi="7119
\mathchardef\rho="711A
\mathchardef\sigma="711B
\mathchardef\tau="711C
\mathchardef\upsilon="711D
\mathchardef\phi="711E
\mathchardef\chi="711F
\mathchardef\psi="7120
\mathchardef\omega="7121
\mathchardef\varepsilon="7122
\mathchardef\vartheta="7123
\mathchardef\varpi="7124
\mathchardef\varrho="7125
\mathchardef\varsigma="7126
\mathchardef\varphi="7127
\mathchardef\imath="717B
\mathchardef\jmath="717C
\mathchardef\ell="7160
\mathchardef\wp="717D
\mathchardef\partial="7140
\mathchardef\flat="715B
\mathchardef\natural="715C
\mathchardef\sharp="715D


\def\err@badsizechange{%
	\immediate\write16{--> Size change not allowed in math mode, ignored}}

\baselinestretch=1000
\tenpoint

\catcode`\@=12					

\twelvepoint

\vsize=8.5in
\hsize=6in

\overfullrule=0pt
\rightline{IASSNS-HEP-95/54}
\rightline{hep-ph/9506458}
\rightline{June 1995}
\vfill
\centerline{\rmb THE ROLE OF GRAVITY IN DETERMINING PHYSICS}
\bigskip
\centerline{{\rmb AT HIGH AS WELL AS LOW ENERGIES}\footnote{$^\dagger$}
{\singlespace Talk presented at the Symposium on ``Physics at the
Planck
Scale,'' held at Puri, India (December, 1994);
to appear in the Proceedings.  This paper overlaps significantly
with the paper ``Bose-Fermi Symmetry:  A Crucial Element in Achieving
Unification,'' IASSNS-HEP-95/26; hep-ph/9506211.}}
\vfill
\centerline{\bf Jogesh C. Pati}
\vskip 0.25 truein
\centerline{\bf Institute for Advanced Study}
\centerline{\bf Princeton, NJ 08540 \ USA}
\centerline{\bf and}
\centerline{{\bf Department of Physics,  University of
Maryland}\footnote{$^*$}{ Address after July 1, 1995.  E-mail:
``pati@umdhep.umd.edu''}}
\centerline{\bf College Park, MD  \ 20742 \ USA}
\vfill

\centerline{\bf Abstract}

It is noted that in the context of a supersymmetric preonic approach
to unification, gravity, though weak, can play an essential role in
determining some crucial aspects of low-energy physics.  These
include:  (i) SUSY-breaking, (ii) electroweak symmetry-breaking, (iii)
generation of masses of quarks and leptons, all of which would vanish
if we turn off gravity.  Such a role of gravity has its roots in the
Witten index theorem which would forbid SUSY-breaking, within the
class of theories under consideration, in the absence of gravity.

\vfill\eject

\item{\rmb I.}~~{\rmb A Prelude}

That gravity should play a major role in determining the nature of
physics near the Planck scale (which is the subject of this meeting)
is of course to be expected because at that scale quantum gravity
becomes supremely dominant.  But traditionally gravity is regarded as
too weak to be relevant to our understanding of microscopic physics
at sub-Planck energies $Q \ltorder M_{Planck}/100$, say) -- that is at
distance scales $\gtorder 10^{-31} cm$.  One main purpose of this talk
is to show that in the context of a supersymmetric preonic approach to
unification, which has recently evolved into a viable and economical
form [1,2,3,4], gravity plays an {\it essential role} in determining
some important aspects of low-energy physics as well.  These include
(i) supersymmetry breaking, (ii) electroweak symmetry breaking, and
(iii) generation of masses of quarks and leptons.

Gravity enters into the game at low energies in this approach as
follows.  The supersymmetric preonic metacolor force becomes strong at
an intermediate scale $\Lambda_M \sim 10^{11}$ GeV.  Owing to the
constraints of the Witten index theorem [5] which forbids
SUSY-breaking in the absence of gravity, the metacolor force by itself
can not break supersymmetry.  It {\it needs} the collaboration of
gravity to induce SUSY-breaking.  At such ``low'' energies ($Q\sim
10^{11}$ GeV), gravity is of course weak and perturbative.  Even then
it induces a negative (mass)$^2$ proportional to Newton's constant for
certain composite scalars which would otherwise be massless in the
limit of SUSY, and thereby induces a VeV for such scalars and in turn
SUSY-breaking.  As a result, the square root of the gravitational
coupling -- i.e., the inverse of $M_{Planck}$ -- enters into the
SUSY-breaking preonic condensates such as $\langle
\bar\psi\psi\rangle$.  These condensates induce mass-splittings
$(\delta m_s)$ between SUSY-partners and even electroweak symmetry
breaking (and thereby $m_W$) as well as $m_q$, all of which thus
become proportional to the much lower scale -- i.e.,
$\Lambda_M(\Lambda_M/M_{Pl})\sim 1~TeV \ll \Lambda_M$ [2].  {\it In this
way, gravity plays the role of a major actor [32], not just that of a
spectator, in determining some crucial elements of low-energy
physics.}  Furthermore, it turns out that the damping by $M_{Planck}$
(noted above) relative to $\Lambda_M$, together with symmetries of the
preonic theory, naturally explains not only why $\delta m_s,~m_W$ and
$m_t$ are so much smaller than $M_{Planck}$ but also why $m_e$ and
even $m_\nu$ are smaller still by many orders of magnitude compared to
$m_t$ [4,1].  In short, the interplay of the metacolor force and
gravity and the symmetries of the SUSY preonic theory turn out to
explain the entire panorama of scales from $M_{Planck}$ to $m_t\sim
m_W$ to $m_e$ to $m_\nu$.

To bring out this role of gravity and to provide motivations for the
underlying preonic approach, I need to say a few words about the
puzzles in particle physics which confront us in the context of the
standard model and the unifying ideas which have been proposed to
resolve some of these puzzles.
\medskip
{\rmb II.}~~
{\rmb Going Beyond the Standard Model}

The standard model of particle physics comprising electroweak and QCD
components has brought a good deal of
synthesis in our understanding of the basic forces of nature, especially
in comparison to its predecessors, and has turned out to be brilliantly
successful in terms of its agreement with experiments.  Yet, as
recognized for some time [1], it falls short as a fundamental theory
because it introduces some 19 parameters.  And it
does not explain (i) family replication;
(ii) the coexistence of the two kinds of matter:  quarks {\it
and} leptons; (iii) the coexistence of the electroweak {\it and} the
QCD forces with their hierarchical strengths $g_1 \ll g_2 \ll
g_3$, as observed at low energies; (iv) quantization of electric charge;
(v) inter and intrafamily mass-hierarchies - {\it i.e.},
$m_{u, d, e} \ll m_{c,s,\mu}\ll m_{t,b,\tau}$ and $m_b\ll m_t$, etc. -
reflected by ratios such as $(m_u/m_t) \sim 10^{-4},~(m_c/m_t)\sim
10^{-2}$ and $(m_b/m_t)\sim {1\over 35}$; and (vi) the origin of diverse
mass scales that span over more than 27 orders of magnitude from
$M_{Planck}$ to $m_W$ to $m_e$ to $m_\nu$, whose ratios involve
very small numbers such as$(m_W/M_{Pl})\sim
10^{-17},~(m_e/M_{Pl})\sim 10^{-22}$ and $(m_\nu/M_{Pl})< 10^{-27}$.
There are in addition the two most basic questions:  (vii) how does
gravity fit into the whole scheme, especially in the context of a good
quantum theory?, and (viii) why is the cosmological constant so small or
zero?
Furthermore, turning to issues in cosmology, it is still a challenge
to obtain a satisfactory particle-physics derived model for both inflation
and baryogenesis.

These issues constitute at present some of the major puzzles of particle
physics and provide motivations for contemplating new
physics beyond the standard model which should shed light on them.
The ideas which have been proposed and which do show promise to
resolve at least some of these puzzles include the following
hypotheses:
\medskip
(1)~~{\bf Grand Unification}: ~~ The hypothesis of grand
unification, which proposes an underlying unity of the fundamental
particles and their forces [6,7,8],
appears attractive
because it explains at once (i) the quantization of electric charge, (ii) the
existence of quarks {\it and} leptons with $Q_e=-Q_p$, and (iii) the
existence of the strong, the electromagnetic and the weak forces with
$g_3\gg g_2\gg g_1$ at low energies, but $g_3=g_2=g_1$ at high energies.
These are among the puzzles listed above and grand unification
resolves all three.
{\it Therefore I believe that the central concept of
grand unification is, very likely, a step in the right direction.}  By
itself, it does not address, however, the remaining
puzzles listed above, including the issues of family replication and
origin of mass-hierarchies.
\medskip
(2)~~{\bf Supersymmetry}:~~ As mentioned before, this is the symmetry that
relates fermions to bosons [9].  As a local
symmetry, it is attractive because it implies the existence of gravity.
It has
the additional virtue that it
helps maintain a large hierarchy in mass-ratios such as
$(m_{\phi}/M_U) \sim 10^{-14}$ and $(m_{\phi}/M_{Pl}) \sim 10^{-17}$,
without
the need for fine tuning, provided, however, such ratios are put in by
hand.
Thus it provides a technical resolution of the gauge hierarchy problem,
{\it but
by itself does not explain the origin of the large hierarchies}.
\medskip
(3)~~{\bf Compositeness}:~~  Here there are {\it two
distinct suggestions}:
\smallskip
(a)~~\undertext{Technicolor}:  The idea of technicolor [10]
proposes that the Higgs bosons are composite but quarks and leptons are
still elementary.  Despite the attractive feature of dynamical symmetry
breaking which eliminates elementary Higgs bosons and thereby the arbitrary
parameters which go with them, this idea
is excluded, at least in its simpler versions, owing to conflicts with
flavor-changing neutral current processes and oblique electroweak
corrections.  The so-called walking technicolor models may be arranged
to avoid some of these conflicts at the expense, however, of excessive
proliferation in elementary constituents.  Furthermore, as a generic
feature, none of these models seem capable of addressing any of the
basic issues listed above, including those of family replication and
fermion mass-hierarchies.  Nor do they go well with the
hypothesis of a unity of the basic forces.
\smallskip
(b)~~\undertext{ Preons}:  By contrast, the idea of preonic
compositeness which proposes that not just the Higgs bosons but also the
quarks and the leptons are composites of a {\it common} set of
constituents called ``preons'' seems much more promising.  Utilizing
supersymmetry to its advantage, the preonic approach has evolved over
the last few years to acquire a form [1-4] which is (a) far more
economical in field-content and especially in parameters than either the
technicolor or the conventional grand unification models, and, (b)
is viable.  Most important, utilizing primarily the symmetries of the
theory (rather than detailed dynamics) and the peculiarities of SUSY QCD
as regards forbiddeness of SUSY-breaking, in the absence of gravity,
the preonic
approach provides simple explanations for the desired protection of
composite quark-lepton masses and at the same time for the origins of
family-replication, inter-family mass-hierarchy and diverse mass scales.
It also provides several testable predictions.  In this sense, though
still unconventional, the preonic approach shows promise in being able
to address certain fundamental issues.  I will return to it shortly.
\medskip
(4)~~{\bf Superstrings}: ~~Last but not least, the idea of
superstrings [11] proposes that the elementary entities are not truly
pointlike but are extended stringlike objects with sizes $\sim
(M_{Planck})^{-1}
\sim 10^{-33}$ cm.  Strings as a rule smoothe out singularities of
point-paritcle field theories.
These theories (which may ultimately be just one) appear
to be most promising in providing a
unified
theory of all matter (spins 0, 1/2, 1, 3/2, 2, ...) and
all the forces of nature including gravity.  Furthermore, by smoothing out
singularities, as mentioned above, they seem capable of
yielding a
well-behaved
quantum theory of gravity.  In principle, assuming that quarks, leptons
and Higgs bosons are elementary, a suitable superstring theory
could also account for the origin of the three families and the Higgs
bosons at the string unification scale, as well as explain all the
parameters of the standard model.  But in practice, this has not happened as
yet.
Some general stumbling blocks of string theories are associated with
the problems of (i) a choice of the ground state (the vacuum)
from among the many solutions and (ii) understanding supersymmetry breaking.

The ideas listed above are, of course, not mutually exclusive.  In fact the
superstring theories already comprise the idea of local
supersymmetry and the central
idea of grand unification.  It remains to be seen, however, whether
they give rise, in accord with the standard belief, to elementary
quarks and leptons, or alternatively to a set of substructure fields --
the preons.  In the following, I first
recall the status of conventional grand unification,
and then provide
a perspective as well as motivations for an alternative approach to
grand unification, based on the idea of preons.  In this case, I
discuss the origin of diverse mass-scales -- from $M_{Planck}$ to
$m_\nu$ -- through the interplay of the metacolor and gravitational
forces.  This brings out the role of gravity in determining some
crucial parameters of low-energy physics which is the main purpose of
this talk.  In the last section, I provide
a summary and a perspective.
\bigskip
\noindent
{\rmb III.}~~{\rmb Grand Unification in the Conventional Approach and
Supersymmetry}
\smallskip

By ``Conventional approach'' to grand unification I mean the one in
which quarks and leptons -- and traditionally the Higgs bosons as well
-- are assumed to be elementary [6,7,8].  Within this approach, there
are two distinct routes to higher unification:  (i) the SU(4)-color
route [6] and (ii) SU(5) [7].
Insisting on a compelling reason
for charge -- quantization, the former naturally introduces the
left-right symmetric gauge structure $G_{224}=SU(2)_L \times
SU(2)_R\times SU(4)_{L+R}^C$ [6], which in turn may be embedded in
anomaly-free simple groups like SO(10) or $E_6$ [12].

It has been known for sometime that the dedicated proton decay
searches at the IMB and the Kamiokande detectors [13], and more
recently the precision measurements of the standard model coupling constants
(in
particular ${\rm sin}^2 \hat\theta_W$) at LEP [14] put severe constraints on
grand unification models without supersymmetry.  Owing to such
constraints, the non-SUSY minimal SU(5) and, for similar reasons, the
one-step breaking non-SUSY SO(10)-model, as well, are now excluded
beyond a shadow of doubt.

But the idea of the union of the coupling constants $g_1, g_2,$ and
$g_3$ can well materialize in accord with the LEP data, if one invokes
supersymmetry [15,16,17] into minimal SU(5) or SO(10).  See Fig. 1, which
shows the {\it impressive meeting} of the three coupling constants of
the minimal supersymmetric standard model (MSSM) with an assumed
SUSY-threshold around 1 TeV.  Such a model can, of course, be embedded
within a minimal SUSY SU(5) or SO(10) model, which would provide the
rationale for the meeting of the coupling constants at a scale $M_U
\approx 2 \times 10^{16}$ GeV, and for their staying together beyond
that scale.

The fact that the coupling constants meet in the context of these models
is reflected by the excellent agreement of their predicted value of
$[{\rm sin}^2 \hat\theta_W (m_z)_{theory}= .2325\pm .005$ (using
$\alpha_s(m_z)=\cdot 12\pm \cdot 01)$ with that determined at LEP:
$[{\rm sin}^2\hat\theta_W(m_z)]_{expt.}= .2316\pm .0003$.
In SUSY SU(5) or SO(10), dimension 5 operators do in general pose
problems for proton decay.  But the relevant parameters of the
SUSY-space can be arranged
to avoid conflict with experiments [18].  The SUSY-extensions of SU(5) or
SO(10) typically lead to prominent strange particle decay modes, e.g.,
$p\to \bar\nu K^+$ and $n\to \bar\nu K^0$, while a 2-step breaking of
SO(10) via the intermediate symmetry $G_{224}$ can also lead to
prominent $\Delta (B-L)=-2$ decay modes of the nucleon via Higgs
exchanges such as $p\to e^-\pi^+\pi^+$ and $n\to e^-\pi^+$ and even
$n\to e^-e^+\nu_e$, etc. in addition to the canonical $e^+\pi^0$-mode
[19].

It is encouraging that the super-Kamiokande (to be completed in April
1996) is expected to be sensitive to the $e^+\pi^0$ mode up to partial
lifetimes of few $\times 10^{34}$ years, to the $\bar\nu K^+$ and
$\bar\nu K^0$ modes with partial lifetimes $\leq 10^{34}$ years and to
the non-canonical $n\to e^-e^+\nu_e$ and $p\to e^-\pi^+\pi^+$ modes with
partial lifetimes $< 10^{33}$ years.  Thus the super-Kamiokande,
together with other forthcoming
facilities, in particular, ICARUS,
provide a {\it big
ray of hope} that first of all one will be able to probe much deeper
into neutrino physics in the near future and second proton-decay may
even be discovered within the twentieth century.
\medskip
\noindent
{\bf Questioning the Conventional Approach}

Focusing attention on the meeting of the coupling constants (Fig. 1),
the question arises:  To what extent does this meeting  reflect the
``truth'' or
is it somehow deceptive?  There are two reasons why such a question is
in order.
\smallskip
(1)~~ First, the unity of forces reflected by the meeting of the
coupling constants in SUSY SU(5) or SO(10) is truly incomplete,
because it comprises only the gauge forces, but not the
Higgs-exchange forces.  {\it The latter are still governed by many arbitrary
parameters -- i.e., the masses, the quartic and the Yukawa couplings of
the Higgs bosons  -- and are thus ununified.}  Such arbitrariness goes
against the central spirit of grand unification and has been the main
reason in my mind since the 1970's (barring an important caveat due to
the growth of superstring theories in the 1980's, see below) to consider
seriously the possibility that the Higgses as well as the quarks and the
leptons are composite.  Furthermore, neither SUSY
SU(5) nor SUSY SO(10),
by itself, has the scope of explaining the origins of (a) the three
families, (b) inter- and intra-family mass-splittings and (c) the
hierarchical mass-scales:  from $M_{Planck}$ to $m_\nu$.
\smallskip
(2)~~ The second reason for questioning the conventional approach is
this:  one might have hoped that one of the two schemes -- i.e.,
the minimal SUSY SU(5) or the SUSY SO(10)-model, or a broken ``grand
unified'' symmetry with relations between its gauge couplings near the
string scale, would emerge from one of
the solutions of the superstring theories [11,20], which
would yield not
only the desired
spectrum of quarks, leptons and Higgs bosons but also just the right
parameters for the Higgs masses as well as their quartic and Yukawa
couplings.  While it seems highly nontrivial that so
many widely varying parameters should come out in just the right way
simply from topological and other constraints of string theories,
it would of course be most
remarkable if that did happen.  {\it But so far it has not}.  There are in
fact a very large number of classically allowed degenerate 4D solutions of the
superstring theories (Calabi-Yau, orbifold and free fermionic, etc.),
although one is not yet able
to choose between them.  Notwithstanding this general difficulty of a
choice, it
is interesting that there are at least some
three-family solutions.  However, not
a single one of these has yielded {\it either} a SUSY SU(5) or an
SO(10)-symmetry, {\it or} a broken ``grand unified'' symmetry involving
direct product of groups, with the desired spectrum
{\it and} Higgs-sector parameters, so as to explain the bizarre pattern of
fermion masses
and mixings of the three families [21].  Note that for a string theory
to yield elementary quarks, leptons and Higgs bosons, either the {\it
entire package} of calculable Higgs-sector parameters, which describe
the masses of
all the fermions and their mixings (subject to perturbative
renormalization), should come out just right, or else the corresponding
solution must be discarded.  This no doubt is a {\it heavy burden}.
For the case of the broken grand unified models, there is
the additional difficulty that the grand unification scale of $2\times
10^{16}~GeV$ obtained from low-energy extrapolation does not match
the string unification scale of about $4\times
10^{17}~GeV$ [22].

Thus, even if a certain superstring theory is the right starting point,
and I believe it is, it is not at all clear, especially in view of the
difficulties mentioned above, that it makes contact with the low-energy
world by yielding elementary quarks, leptons and Higgs bosons.  In this
sense, it seems prudent to keep open the possibility that the meeting of
the coupling constants in the context of conventional grand unification,
which after all corresponds to predicting just one number -- i.e.,
$sin^2\theta_W$ -- correctly, may be fortuitous.  Such a meeting should
at least be
viewed with caution as regards inferring the extent to which it
reflects the ``truth'' because there are in fact alternative ways by
which such a meeting can occur (see discussions below).
\medskip
\noindent
{\rmb IV.}~~{\rmb The Preonic Approach to Unification and Supersymmetry}

This brings me to consider an alternative approach to unification based
on the ideas of preons and local supersymmetry [1-4].  Although
the general idea of preons is old [23], the particular approach [1-4]
which I am about to present has evolved in the last few years.  It is
still unconventional,  despite its promising features.  Its
lagrangian introduces only six positive and six negative chiral preonic
superfields which define the two flavor and four color attributes of a
quark-lepton family and possess only {\it the minimal gauge
interactions} corresponding to flavor-color and metacolor gauge
symmetries [6].  But the lagrangian {\it is devoid altogether of the
Higgs sector since its superpotential is zero owing to gauge and
non-anomalous R-symmetry}.  Therefore, it is free from all the arbitrary
Higgs-mass, quartic and Yukawa coupling parameters which arise in the
conventional approach to grand unification.  This brings {\it real
economy}.  {\it In fact, the preon model possesses just three (or four)
gauge
coupling parameters which are the only parameters of the model and even
these few would merge into one near the Planck scale if there is an
underlying unity of forces as we envisage} [24].  By contrast, the
standard model has 19 and conventional SUSY grand unification models
have over 15 parameters.  As mentioned in the introductory chapter, in
addition to economy, the main motivations for pursuing the preonic
approach are that it provides simple explanations for (a) the protection
of the masses of the composite quarks and leptons [2], (b) family
replication [3], (c) inter-family mass-hierarchy $(m_{u,d,e}\ll
m_{c,s,\mu}\ll m_{t,b,\pi})$ [4], and (d) diverse mass-scales [1].  At
the same time, it is viable with respect to observed processes including
flavor-changing neutral current processes (see remarks later) and
oblique electroweak corrections.

Fermion-boson partnership in a SUSY theory,
(i.e. $\psi
\leftrightarrow \varphi$ and $v_{\mu} \leftrightarrow \lambda$ or
$\overline{\lambda}$ etc.), leads to several alternative three-particle
combinations with identical quantum numbers, which can make a left-chiral
$SU(2)_L$-doublet family $q^i_L$ -- e.g. (i)~$\sigma_{\mu \nu} \psi^f_L
\varphi^{c^*}_R v_{\mu \nu}$, (ii)~$\sigma^{\mu \nu} \varphi^f_L
\psi^{c^*}_R
v_{\mu \nu}$, (iii)~$\psi^f_L \psi^{c^*}_R \lambda$ and (iv)~$\varphi^f_L
(\sigma^{\mu} \overline{\lambda}) \partial_{\mu} \varphi^{c^*}_R$.  Here
$f$
and $c$ denote flavor and color quantum numbers.
{\it The plurality of these combinations, which stems
because of SUSY, is in essence the origin of family-replication}.
By constructing composite superfields, Babu, Stremnitzer and I
showed [3] that at the level of minimum dimensional composite
operators
(somewhat analogous to $qqq$ for QCD) there are just three linearly
independent
chiral families $q^i_{L,R}$, and, in addition, two {\it vector-like
families}
$Q_{L,R}$ and $Q^{\prime}_{L,R}$, which couple vectorially to $W_L$'s and
$W_R$'s respectively.  Each of these composite families with spin-1/2 is,
of course,
accompanied by its scalar superpartner.  {\it We thus see that one
good answer
to Rabi's famous question:  ``Who ordered that?'', is supersymmetry and
compositeness.}

Certain novel features in the dynamics of a class of SUSY QCD theories,
in particular (as mentioned in the introduction) the forbidding of
SUSY-breaking in the absence of gravity [5,2], and symmetries of the
underlying preonic theory, play crucial roles in obtaining the other
desired results -- (a), (c) and (d), mentioned above.  The reader is
referred to the papers in Refs. 1-4 and in particular to a recent
review of the preonic approach in Ref. 25 for details of the two broad
dynamical assumptions [26] and the reasons underlying a derivation of these
results.  One attractive feature of the model, which emerges primarily
through the symmetries of the underlying lagrangian, is that the two
vector-like families $Q_{L,R}$ and $Q_{L,R}^\prime$ (mentioned above)
acquire masses of order 1 TeV, while the three chiral families acquire
their masses primarily through their spontaneously induced mixings with
the two vector-like families.  This feature automatically explains why
the electron family is so light compared to the tau-family and (owing to
additional symmetries) why the masses of the muon-family lie
intermediate between those of the electron and the tau-families.  In
particular, the model explains why $m_e\sim 1~MeV$ while $m_t\approx
100-180~GeV$, i.e., why $(m_e/m_t)\sim 10^{-5}$.

It is shown [1,4] that
the model is
capable of generating all the diverse scales -- from $M_{Planck}$ to
$m_{\nu}$
-- and thereby the small numbers such as $(m_W/M_{Pl}) \sim (m_t/M_{Pl})
\sim
10^{-17}, (m_C/M_{Pl}) \sim 10^{-19}, (m_e/M_{Pl}) \sim 10^{-22}$, and
$(m_{\nu}/M_{Pl}) < 10^{-27}$ -- {\it in terms of just one fundamental
input
parameter}: the coupling constant $\alpha_M$ associated with the metacolor
force.   This comes about as follows.
Corresponding to an input value $\overline{\alpha}_M \approx 1/27$
to
1/32 at $M_{Pl}/10$, the metacolor force generated by $SU(N)_M$ becomes
strong
at a scale $\Lambda_M \approx 10^{11} GeV$ for $N$=5 to 6.  Thus the first
big
step in the hierarchical ladder leading to the small number
$(\Lambda_M/M_{Pl})
\sim 10^{-8}$ arises naturally through renormalization group equations due
to
the slow logarithmic growth of $\overline{\alpha}_M$ and its perturbative
input
value at $M_{Pl}/10$.

The next step arises due to the constraint on SUSY breaking, which is
forbidden [5], except for the presence of gravity.  As mentioned
before in section I, SUSY-breaking condensates like
$\langle\lambda\lambda\rangle$
and $\langle\bar\psi \psi\rangle$ are thus naturally damped by
$(\Lambda_M/M_{Pl})$ [2].  These induce (a) SUSY-breaking
mass-splittings
$\delta m_S\sim {\cal O}(\Lambda_M(\Lambda_M/M_{Pl}))\sim{\cal O}(1~TeV)$
and (b) $m_W\sim m_t\sim (1/10){\cal O}(\Lambda_M(\Lambda_M/M_{Pl}))$
$\sim {\cal O}(100~GeV)$.  Note the natural origin of the small numbers:
$(\delta m_s/M_{Pl})\sim 10^{-16}$ and $(m_W/M_{Pl})\sim 10^{-17}$.  As
also noted above, symmetries of the $5\times 5$ fermion mass-matrix take
us down to still lower scales -- in particular to $m_e\sim {\cal O}
(1~MeV)$, thus accounting for the tiny number $(m_e/M_{Pl}) \sim
10^{-22}$.

Finally, the familiar see-saw mechanism for neutrinos with
$m(\nu_R^i)\sim\Lambda_M\sim 10^{11}~GeV$ and $m(\nu^i)_{Dirac}\propto
\Lambda_M(\Lambda_M/M_{Pl})$ yields $m(\nu_L^i)\leq
10^{-3}~M_{Pl}(\Lambda_M/M_{Pl})^3\sim 10^{-27}M_{Pl}$.  In this way,
the model provides a {\it common origin} of all the diverse mass scales
-- from $M_{Pl}$ to $m_\nu$, and of the associated small numbers, as
desired [1].  {\it This constitutes a unification of scales which is
fundamentally as important as the unification of forces.}  By and by
we see, as indicated in the beginning, that gravity plays an essential
role in determining some crucial parameters of low-energy physicss,
such as $\delta m_s, m_W, m_t, m_e$ and $m_\nu$, all of which will
vanish if we turn off gravity.

Furthermore, using the values of the standard model gauge couplings
measured at LEP and the spectrum of the preon model above and below the
preon-binding scale $\Lambda_M \sim 10^{11}~GeV$, it is found (see Fig.
2) that the flavor-color gauge symmetry being
$SU(2)_L\times U(1)_R\times SU(4)^c$ near the Planck scale and the
metacolor gauge symmetry being either $SU(5)$ [24] or $SU(6)$ [27], the
gauge couplings do tend to meet near the Planck scale.  This opens up a
novel possibility for grand unification at the preon level and thereby a
possible new route for superstring theories to make connection with the
low-energy world.

Last but not least the preon model leads to some {\it crucial
predictions} which include the existence of the two vector-like families
at the TeV-scale.  [See Refs. 1,4 and 25 for a list of predictions.]
These two families can be searched for at the forthcoming LHC, the
$e^-e^+$ next linear collider (in planning) and especially at a future
version of the now-extinct SSC.  Their discovery or non-discovery with
masses up to few TeV will clearly vindicate or exclude the preonic
approach developed in Refs. 1-4.
\bigskip
\noindent
{\rmb V.~~Summary and a Perspective}

The passage from the standard model to grand unification to
supersymmetry and superstrings generates rightfully the
hope for achieving an ultimate synthesis of all matter and its forces.
The ideas and principles underlying this passage are those of:
\smallskip
$\bullet$~~~~~Local gauge invariance,
\smallskip
$\bullet$~~~~~Spontaneous breaking of symmetries through either elementary
or composite Higgs boson,
\smallskip
$\bullet$~~~~~Supersymmetry and
\smallskip
$\bullet$~~~~~Extended string-like rather than point-like elementary
entities.
\smallskip
\noindent
Of these, the relevance of the first two ideas to nature -- i.e., local
gauge invariance and spontaneous symmetry breaking -- is amply
demonstrated by the success of the standard model comprising electroweak
and QCD forces.  Even then, the precise origin of electroweak symmetry
breaking -- i.e., whether it occurs through the vacuum expectation value
of an elementary or a composite Higgs boson -- is still not clear.

As explained above, although unconventional, the preonic alternative,
which proposes that the Higgs bosons as well as the quarks and the
leptons are composite, seems to be viable and deserves serious
consideration. Its drawback at present is that it relies
on two dynamical assumptions [26] as regards (a) confinement and
(b) the pattern of symmetry-breaking that might occur at the preonic
metacolor scale of $10^{11}$ GeV.  While these two assumptions are
not implausible, they have not yet been proven [28].  Not
withstanding this drawback, the preonic approach has the clear advantage
that it is the most economical model around.  Furthermore, the simplicity
with which it explains the origin of inter-family mass-hierarchy and of
the diverse mass-scales (from $M_{Planck}$ to $m_W$ to $m_\nu$) lends
support to this approach.  As noted above, the preonic approach also
retains the central spirit of grand unification as regards the meeting
of the coupling constants.  Its major strength is that it offers some
crucial predictions, in particular the existence of two vector-like
families at the TeV-scale (see above), by which it can be falsified or
vindicated.  For these reasons, it seems
prudent to keep an open mind about the prospects of both the conventional
as well as the preonic alternative.

Turning now to the relevance of supersymmetry to nature, although it is
yet to show in experiments, just by uniting bosons and fermions it seems
to play an {\it essential role} in every attempt at higher unification,
beyond that of the standard model.  These include:  (i) the
conventional approach to grand unification, (ii) the preonic approach,
and (iii) superstrings.

Turning finally to the relevance of superstring theories to nature,
motivations for these theories at present are entirely
theoretical, somewhat analogous to but considerably beyond those for only
supersymmetry.
As mentioned before, the superstring theories provide the scope for the
greatest synthesis so far in particle physics in that they seem capable
of unifying {\it all matter} (spins 0, 1/2, 3/2, 2 and higher)
as vibrational modes
of the string and also {\it all their interactions}, which include not
only the gauge forces and gravity but also the apparently non-gauge
Higgs-type Yukawa and quartic couplings, within a single coherent
framework.  {\it The most attractive feature is that the superstring theories
permit no dimensionless parameter at the fundamental level}.  Equally
important, they provide the scope for yielding a good quantum theory of
gravity.

For these reasons, I believe that superstring theories possess many (or
most) of the crucial ingredients of a ``final theory'' --
``the theory of
everything''.  {\it But I also believe that, as they stand, they do not
constitute the whole of an ultimate theory}, because, first and foremost,
in spite of the desirable feature that they constrain the gauge symmetry,
the spectrum and
the S-matrix elements (interactions), they are not generated by an
underlying principle analogous to that of general coordinate or gauge
invariance.  Second, as a practical matter, they do not yet explain why we live
in $3+1$ dimensions, and given the fact that supersymmetry does break in
the real world, they do not explain why the cosmological constant is so
small or zero.  Third, they also do not yet provide a consistent understanding
of (a) supersymmetry breaking and (b) choice of the ground state.
Resolutions of some or all of these latter issues, which may well be
inter-related, would clearly involve an understanding of the
non-perturbative aspects and the symmetries of superstring dynamics.
Recent developments which include the ideas of duality symmetries [29]
and the realization that the strong-coupling limit of certain
superstring theories is equivalent to the weak-coupling limit of certain
other theories [30], permitting the elegant and bold conjecture [31] that
there is just one superstring theory, may evolve into a form so as to
achieve the lofty goal of solving superstring dynamics.  It remains to
be seen, however, as to how much of the resolution of the issues
mentioned above could come ``merely'' from our understanding of the
non-perturbative dynamics of the existing string theories and how much of such
a resolution would involve altogether new ingredients beyond the
framework of existing string theories, which may call for some {\it
radical changes} in our concepts at a fundamental level.

As another practical matter, for reasons mentioned in Sections III and
IV, it is
far from clear that the superstring theories make connections with the
low-energy world by yielding elementary quarks, leptons and Higgs
bosons.  The preonic approach, though unconventional, provides a viable
and attractive alternative to the conventional approach.  It therefore
remains to be seen whether the right superstring theory would yield the
elementary quark-lepton-Higgs system with the entire ``right package''
of Higgs-sector parameters or, instead, the preonic spectrum and the
associated gauge symmetry.  In the latter case, the superstring theory
would, of course, be relieved from yielding the right package of such
Higgs sector-parameters because the Higgs-sector is simply absent in
the preonic theory.

One last remark, our understanding of superstring theories is rather
premature.  It would clearly take some time -- optimistically a decade
but conservatively several decades --for us to understand (and
this may be optimistic) the true nature of superstring theories and to
discover the missing ingredients (alluded to above) in these theories,
which
together would help resolve the issues mentioned above.  Meanwhile,
regardless of these developments in the future, supersymmetry has
clearly evolved as a great synthesizing principle.  It is a common
denominator and a central feature in all the attempts at higher
unification which I mentioned above.
{\it As such, it is hard to imagine how nature could
have formulated her laws without the aid of supersymmetry}.
Fortunately, unlike some other
concepts, the relevance of supersymmetry to particle physics, as commonly
conceived, can
be established or falsified, depending upon whether the superpartners are
discovered with masses in the range of 100 GeV to a few TeV
or found to be absent in the forthcoming accelerators.

To conclude, the point of view brought forth in this talk is this:   in
the context of a supersymmetric preonic approach to unification, weak
perturbative gravity, in collaboration with the preonic metacolor
force, can play an {\it active role} in determining some crucial
aspects
of low-energy physics.  Such an interplay between these two forces
would in particular permit us to resolve one of the major puzzles in
particle physics pertaining to the origin of diverse mass-scales that
span over more than 27 orders of magnitude -- from $M_{Planck}$ to
$m_W$ to $m_e$ to $m_\nu$.  By linking these diverse mass scales, one
obtains a unification of scales [1] which is fundamentally as
important as the unification of forces.  This attractive scenario that
emphasizes the active role of gravity at low energies can of course be
realized, as far as I can see, only in the context of supersymmetry
and preons [32].  Fortunately, just like supersymmetry, the preonic
approach provides some crucial tests, in particular the existence of
the two vector-like families with masses of order 1 TeV, which can be
searched, together with SUSY particles and the Higgses, at the LHC,
$e^+e^-$ NLC and a future version of the now-extinct SSC.  It is only
these experimental facilities which can ultimately free us from the
present bottleneck in particle physics and hopefully tell us which of
our preconceived notions about elementary particles are right, if any,
and which are wrong.
\bigskip
{\rmb VI.~~ Acknowledgements}

The research described in this talk is supported in part by the NSF
grant number 9421387.  This manuscript is prepared during the author's
visit to the Institute for Advanced Study, Princeton, New Jersey, which
is supported in part by a sabbatical leave grant from the University of
Maryland and in part by a grant by the IAS.  It is a pleasure to thank
Keith Dienes, Alon Faraggi and especially Edward Witten for several
discussions which
were helpful in preparing this manuscript.  I wish to thank the
organizers of the conference,  especially Jnana Maharana,
for arranging a stimulating conference in a beautiful setting at the
Puri Beach and
for the kind hospitality.
I would also like to thank Valerie Nowak
for her most generous cooperation in typing this manuscript.
\medskip
\centerline{\rmb References}
\item{[1]}
J.C. Pati, Phys. Lett. {\bf B228} (1989) 228.
\item{[2]}
J.C. Pati, M. Cvetic and H. Sharatchandra, Phys. Rev. Lett. {\bf 58}
(1987) 851.
\item{[3]}
K.S. Babu, J.C. Pati and H. Stremnitzer, Phys. Lett. {\bf B256} (1991)
206.
\item{[4]}
K.S. Babu, J.C. Pati and H. Stremnitzer, Phys. Rev. Lett. {\bf 67}
(1991) 1688.
\item{[5]}
E. Witten, Nucl. Phys. {\bf B185} (1981) 513; {\bf B202} (1983) 253;
E. Cohen and L. Gomez, Phys. Lett. {\bf 52} (1984) 237.
\item{[6]}
J.C. Pati and Abdus Salam; Proc. 15th High Energy Conference, Batavia.,
reported by J.D. Bjorken, Vol. 2, p. 301 (1972); Phys. Rev. {\bf 8} (1973)
1240; Phys. Rev. Lett. {\bf 31} (1973) 661; Phys. Rev. {\bf D10} (1974)
275.
\item{[7]}
H. Georgi and S.L. Glashow, Phys. Rev. Lett. {\bf 52} (1974) 438.
\item{[8]}
H. Georgi, H. Quinn, and S. Weinberg, Phys. Rev. Lett. {\bf 33} (1974)
451.
\item{[9]}
Y.A. Gelfand and E.S. Likhtman, JETP Lett. {\bf 13} (1971) 323; J.
Wess and B. Zumino, Nucl. Phys. {\bf B70} (1974)139; Phys. Lett. {\bf
49B}(1974) 52; D. Volkov and V.P. Akulov, JETP Lett. {\bf 16}
(1972) 438.
\item{[10]}
For a review, see E. Farhi and L. Susskind, Phys. Rev. {\bf 74} (1981)
277 and references therein.
\item{[11]}
M. Green and J. Schwarz, Phys. Lett. {\bf 149B} (1984) 117; D. Gross, J.
Harvey, E. Martinec and R. Rohm, Phys. Rev. Lett {\bf 54} (1985) 502; P.
Candelas, G. Horowitz, A. Strominger and E. Witten, Nucl. Phys. {\bf
B258} (1985) 46.
\item{[12]}
SO(10):  H. Georgi, Proc. AIP Conf. Williamsburg (1994); H. Fritzsch and
P. Minkowski, Ann. Phys. (NY) {\bf 93} (1975) 193.  E(6):  F. Gursey, P.
Ramond and P. Sikivie, Phys. Lett. {\bf 60B} (1976) 177.
\item{[13]}
Particle Data Group, Review of Particle Properties, Phys. Rev. {\bf
D45}, Part II, SI-S584, June 1, 1992.
\item{[14]}
LEP data, Particle Data Group (June, 1994).
\item{[15]}
S. Dimopoulos and H. Georgi, Nucl. Phys. {\bf B193} (1981) 150; N.
Sakai, Z. Phys. {\bf C11} (1982) 153.
\item{[16]}
P. Langacker and M. Luo, Phys. Rev. {\bf D44} (1991) 817; U. Amaldi, W.
de Boer and H. Furstenau, Phys. Lett. {\bf B260} (1991) 447; J. Ellis,
S. Kelley and D.V. Nanopoulos, Phys. Lett. {\bf B260} (1991) 131; F.
Anselmo, L. Cifarelli, A. Peterman and A. Zichichi, Nuov. Cim. {\bf
A104} (1991) 1817.
\item{[17]}
P. Langacker, Review talk at Gatlinburg Conference, June '94,
HEP-PH-9411247.
\item{[18]}
For analysis of this type, see e.g. R. Arnowitt and P. Nath, Phys. Rev.
Lett. {\bf 69} (1992) 725; K. Inoue, M. Kawasaki, M. Yamaguchi and T.
Yanagida, Phys. Rev. {\bf D45} (1992) 328; G.G. Ross and R.G. Roberts,
Nucl. Phys. {\bf B377} (1992) 571; and J.L. Lopez, D. Nanopoulos and H.
Pois, Phys. Rev. Lett. {\bf 47} (1993) 2468.  Other references may be
found in the last paper.
\item{[19]}
J.C. Pati, A. Salam and U. Sarkar, Phys. Lett. {\bf 133B} (1983) 330;
J.C. Pati, Phys. Rev. Rap. Comm. {\bf D29} (1983) 1549.
\item{[20]}
H. Kawai, D. Lewellen and S. Tye, Nucl. Phys. {\bf B288} (1987) 1;
I. Antoniadis, C. Bachas and C. Kounnas, Nucl. Phys. {\bf B289} (1987)
187.
\item{[21]}
Some partially successful three-family solutions with top acquiring a
mass of the right value ($\approx 175~GeV$) and all the other fermions
being massless at the level of cubic Yukawa couplings have been obtained
with $Z_2\times Z_2$ orbifold compactification by A. Faraggi, Phys. Lett.
{\bf B274} (1992) 47; Nucl. Phys. {\bf B416} (1994) 63; and J. Lopez, D.
Nanopoulos and A. Zichichi, Texas A\&M preprint CTP AMU-06/95).
In these attempts,
all the other masses and mixings including $m_e\sim {\cal O}(1~MeV)$ are
attributed to in-principle calculable higher dimensional operators.  It
seems optimistic that the entire package of effective parameters
would come out correctly this way with the desired hierarchy.  But, of
course, there is no argument that they cannot.  Thus it seems most
desirable to pursue this approach as far as one can.  This is why I
personally keep
an open mind with regard to both the conventional approach and the
preonic alternative.
\item{[22]}
V.S. Kaplunovsky, Nucl. Phys. {\bf B307} (1988) 145; recently, K.R. Dienes
and A.E. Faraggi (preprints hep-th/9505018 and hep-th/9505046) provide
general arguments why string-threshold corrections arising from the
massive tower of states are naturally suppressed and,
thus, these corrections do not account for such a mismatch between
the two scales.  They and other authors have noted that string theories
tend to give extra matter, which, if they acquire masses in the right
range, could eliminate the mismatch.  For a recent discussion of some
relevant issues pertaining to string-unification, see these two papers
as well as L.E. Ib\'a\~nez, talk at Strings '95, USC, March 1995,
FTUAM 95/15-ReV.
\item{[23]}
Old works on composite models for quarks {\it and} leptons include the
presently-pursued idea of flavon-chromon preons which was introduced
in the paper of J.C.
Pati and A. Salam, Phys. Rev. {\bf D10} (1974) 275 (Footnote 7).  A
similar idea that treated only quarks but not leptons as composite was
considered independently by O.W. Greenberg (private communication to
JCP).  This idea has been subsequently considered by Pati and Salam in a
set of papers (1975-80) and by several other authors -- with W's treated
as composites in some of them -- see e.g., H. Terezawa, Prog. Theor.
Phys. {\bf 64} (1980) 1763, H. Fritzsch and G. Mandelbaum, Phys. Lett.
{\bf 102B}, (1981) 113, and O.W. Greenberg and J. Sucher, Phys. Lett.
{\bf 99B} (1981) 339; and supersymmetric versions in J.C. Pati
and A. Salam, Nucl. Phys. {\bf B214} (1983) 109; {\it ibid.} {\bf B234}
(1984) 223 and by R. Barbieri, Phys. Lett. {\bf 121B} (1983) 43.  None
of these works provided a reason, however, for (a) the protection of
composite quark-lepton masses, (b) family-replication and (c)
inter-family mass-hierarchy.  The interesting idea of quasi-Nambu
Goldstone fermions suggested by W. B\"uchmuller, R. Peccei and T.
Yanagida, Phys. Lett. {\bf 124B} (1983) 67, provided a partial reason
for some of these issues, in particular for (a), but had problems of
internal consistency as regards SUSY-breaking while maintaining the
lightness of composite quarks and generating effective gauge symmetries.
The idea of SUSY-compositeness developed in Refs. 6-9 and Ref. 23
introduces a {\it new phase} in the preonic approach in that (i) it avoids the
problems of technicolor and (ii) it seems capable of incorporating the
idea of grand unification [23], while providing a reason for each of the
issues (a), (b) and (c) mentioned above.  It is these features which
seem to make the new approach a {\it viable alternative} to
the conventional approach to grand unification.
\item{[24]}
K.S. Babu and J.C. Pati, Phys. Rev. Rap. Comm. {\bf D48} (1993) R1921.
\item{[25]}
For a recent review of the preonic approach, see J.C. Pati ``Towards a
Unified Origin of Forces, Families and Mass Scales,''
hep-ph/9505227, to appear in the Proc. of the 1994 Int'l. Conf. on
B-Physics, held at Nagoya, Japan (Oct. 26-29, 1994).
\item{[26]}
For a SUSY preon theory, based on an asymptotically free metacolor
force, generated by an $SU(N_c)$ gauge theory, with $N_f=6$ and $N_c=
N_f$ or $N_f-2$, the two main assumptions [25] are:  (i) the
metacolor force confines preons at a scale $\Lambda_M\sim 10^{11}$
GeV;
(ii) it makes a few preonic condensates and thereby breaks dynamically
the (approximate) global symmetry $G$ of the preonic theory as well
as the flavor-color gauge symmetry (which is a subgroup of $G$) into
just the standard model gauge symmetry at $\Lambda_M$, while
preserving SUSY.
\item{[27]}
The threshold effects at $\Lambda_M$ which permit unity with metacolor
symmetry being SU(6) have been considered by K.S. Babu, J.C. Pati and
M. Parida (to appear).
\item{[28]}
The validity of the two assumptions stated in Ref. 26 in the light of
certain general results derived recently by N. Seiberg [Proc. of
PASCOS '94] is not clear either way if one allows for soft
SUSY-breaking scalar preon (mass)$^2$-terms, which are expected to be
induced by a hidden sector.  There are a few other relevant
differences between the premises of Seiberg's work and those of the
preonic approach which are noted in Ref. 25 [see especially footnote
26 of this paper].
\item{[29]}
See e.g., A. Sen, Int. J. Mod. Phys. {\bf A9} (1994) 3707, hep-th/9402002
and hep-th/9402032; J.H. Schwarz, hep-th/9411178; C.M. Hull and P.K.
Townsend, QMW94-39, R/94/33 and references therein.
\item{[30]}
E. Witten, preprint IASSNS-HEP-95/18, hepth-9503124.
\item{[31]}
E. Witten, Ref. 34 and private communications.
\item{[32]}
Note that the role of gravity in case of the SUSY preonic approach
differs characteristically from that in the conventional approach.
While for the former it enters into the dynamics of the formation of
SUSY-breaking condensates, for the latter it plays the role of a
messenger in transmitting SUSY-breaking effects from the hidden to the
observable sector.
\bye

\centerline{\rmb Figure 2}
\noindent
{\bf A grand fiesta of new physics at ${\bmit 10^{11}~GeV}$:}
The preonic approach suggests the existence of a rich variety of new
physics, listed above, at the $10^{11}~GeV$ scale, all of which
could have a {\it common origin} through one and the same source:  the locally
supersymmetric metacolor force, operating in the observable sector, with
a scale-parameter $\Lambda_M\sim 10^{11}~GeV$.  As the metacolor force
becomes strong at this scale, it generates the set of phenomena,depicted
above, some of
which preserve SUSY (such as the breakings of $SU(4)^c$, B-L and PQ
symmetry) and some which do not (such as $\delta m_s,~m_q$ and $m_W$).
Since SUSY-breaking effects {\it need} the collaboration of gravity
(even though perturbative) with the metacolor force, they are naturally
damped compared to the metacolor scale by the gravitational coupling.
Thus arises the hierarchy of scales -- e.g., $m_W\sim m_t\sim\delta
m_s\sim \Lambda_M(\Lambda_M/M_{Pl})\sim
M_{Pl}(\Lambda_M/M_{Pl})^2\ll M_{Pl}$.

\bye